\documentclass[12pt]{article}
\usepackage{amsfonts}
\usepackage{amssymb}
\usepackage{graphics,psboxit,amsmath}
\usepackage{subfigure}
\usepackage{graphicx}
\usepackage{verbatim}
\usepackage{epic}

\def\hybrid{\topmargin 0pt \oddsidemargin 0pt 
        \headheight 0pt \headsep 0pt
        \textwidth 16,5cm 
        \textheight 23cm 
        \marginparwidth .875in
        \parskip 5pt plus 1pt \jot = 1.5ex}

\hybrid

\catcode`\@=11

\def\marginnote#1{}
\newcount\hour
\newcount\minute
\newtoks\amorpm
\hour=\time\divide\hour by60
\minute=\time{\multiply\hour by60 \global\advance\minute by-\hour}
\edef\standardtime{{\ifnum\hour<12 \global\amorpm={am}%
        \else\global\amorpm={pm}\advance\hour by-12 \fi
        \ifnum\hour=0 \hour=12 \fi
        \number\hour:\ifnum\minute<10 0\fi\number\minute\the\amorpm}}
\edef\militarytime{\number\hour:\ifnum\minute<10 0\fi\number\minute}

\def\draftlabel#1{{\@bsphack\if@filesw {\let\thepage\relax
   \xdef\@gtempa{\write\@auxout{\string
      \newlabel{#1}{{\@currentlabel}{\thepage}}}}}\@gtempa
   \if@nobreak \ifvmode\nobreak\fi\fi\fi\@esphack}
        \gdef\@eqnlabel{#1}}
\def\@eqnlabel{}
\def\@vacuum{}
\def\draftmarginnote#1{\marginpar{\raggedright\scriptsize\tt#1}}

\def\draft{\oddsidemargin -.5truein
        \def\@oddfoot{\sl preliminary draft \hfil
        \rm\thepage\hfil\sl\today\quad\militarytime}
        \let\@evenfoot\@oddfoot \overfullrule 3pt
        \let\label=\draftlabel
        \let\marginnote=\draftmarginnote
   \def\@eqnnum{(\theequation)\rlap{\kern\marginparsep\tt\@eqnlabel}%
\global\let\@eqnlabel\@vacuum} }

\def\draft2{
        \def\@oddfoot{\sl preliminary draft \hfil
        \rm\thepage\hfil\sl\today\quad\militarytime}
        \let\@evenfoot\@oddfoot \overfullrule 3pt
        \let\label=\draftlabel
        \let\marginnote=\draftmarginnote
   \def\@eqnnum{(\theequation)\rlap{\kern\marginparsep\tt\@eqnlabel}%
\global\let\@eqnlabel\@vacuum} }

\def\preprint{\twocolumn\sloppy\flushbottom\parindent 2em
        \leftmargini 2em\leftmarginv .5em\leftmarginvi .5em
        \oddsidemargin -.5in \evensidemargin -.5in
        \columnsep .4in \footheight 0pt
        \textwidth 10.in \topmargin -.4in
        \headheight 12pt \topskip .4in
        \textheight 6.9in \footskip 0pt
        \def\@oddhead{\thepage\hfil\addtocounter{page}{1}\thepage}
        \let\@evenhead\@oddhead \def\@oddfoot{} \def\@evenfoot{} }

\def\numberbysection{\@addtoreset{equation}{section}
        \def\theequation{\thesection.\arabic{equation}}}

\def\underline#1{\relax\ifmmode\@@underline#1\else
        $\@@underline{\hbox{#1}}$\relax\fi}

\def\titlepage{\@restonecolfalse\if@twocolumn\@restonecoltrue\onecolumn
     \else \newpage \fi \thispagestyle{empty}\c@page\z@
        \def\thefootnote{\fnsymbol{footnote}} }

\def\endtitlepage{\if@restonecol\twocolumn \else \newpage \fi
        \def\thefootnote{\arabic{footnote}}
        \setcounter{footnote}{0}} 

\catcode`@=12
\relax

\def\figcap{\section*{Figure Captions\markboth
        {FIGURECAPTIONS}{FIGURECAPTIONS}}\list
        {Figure \arabic{enumi}:\hfill}{\settowidth\labelwidth{Figure
999:}
        \leftmargin\labelwidth
        \advance\leftmargin\labelsep\usecounter{enumi}}}
 \relax
\def\tablecap{\section*{Table Captions\markboth
        {TABLECAPTIONS}{TABLECAPTIONS}}\list
        {Table \arabic{enumi}:\hfill}{\settowidth\labelwidth{Table
999:}
        \leftmargin\labelwidth
        \advance\leftmargin\labelsep\usecounter{enumi}}}
 \relax
\def\reflist{\section*{References\markboth
        {REFLIST}{REFLIST}}\list
        {[\arabic{enumi}]\hfill}{\settowidth\labelwidth{[999]}
        \leftmargin\labelwidth
        \advance\leftmargin\labelsep\usecounter{enumi}}}
 \relax
\makeatletter
\newcounter{pubctr}
\def\publist{\@ifnextchar[{\@publist}{\@@publist}}
\def\@publist[#1]{\list
        {[\arabic{pubctr}]\hfill}{\settowidth\labelwidth{[999]}
        \leftmargin\labelwidth
        \advance\leftmargin\labelsep
        \@nmbrlisttrue\def\@listctr{pubctr}
        \setcounter{pubctr}{#1}\addtocounter{pubctr}{-1}}}
\def\@@publist{\list
        {[\arabic{pubctr}]\hfill}{\settowidth\labelwidth{[999]}
        \leftmargin\labelwidth
        \advance\leftmargin\labelsep
        \@nmbrlisttrue\def\@listctr{pubctr}}}
 \relax
\makeatother

\def\ba{\begin{equation}}
\def\ea{\end{equation}}

\def\no{\noindent}

\def\IR{\relax{\rm I\kern-.18em R}}

\begin{document}

\renewcommand{\theequation}{\thesection.\arabic{equation}}
\csname @addtoreset\endcsname{equation}{section}

\newcommand{\eqn}[1]{(\ref{#1})}
\newcommand{\be}{\begin{eqnarray}}
\newcommand{\ee}{\end{eqnarray}}
\newcommand{\non}{\nonumber}
\begin{titlepage}
\strut\hfill
\begin{center}

\vskip -1 cm


\vskip 2 cm

{ \Large \bf Geons found include non-susy CDM particle
and non-singular ``Kerr-Newman" models}

{\bf Nikolaos A. Batakis}

\vskip 0.2in

Department of Physics, University of Ioannina, \\
45110 Ioannina,  Greece\\
{\footnotesize{\tt (nbatakis@uoi.gr)}}\\

\end{center}

\vskip .4in

\centerline{\bf Abstract}

\no 
The antisoliton-soliton ${\cal G}={\cal S_-}\!\vee{\cal S_+}$ neutral state is a proper geon  in a family of stable 
squashed-$S^3\times\IR$ pp-wave electrovacua along a primordial $Q/r^2$ field. With ${\cal S_-}$ propagating 
backwards in time, the dominant EM field is that of an effective electric-dipole moment ${\sf p}$. If disjointed, 
the ${\cal S_\pm}$  carry $\pm Q$ charge (on a round-$S^2$ {\em physical}  singularity of radius $r_o$) as 
non-singular alternatives to the Kerr-Newman solution. ${\cal G}$ has three scales (gravitational $\kappa$, 
metric scale, NUT-charge $\kappa Q=2r_o$) in a full 4-scale hierarchy without supersymmetry. A particular 
${\cal G}$ with effective mass and a near-zero ${\sf p}$ is proposed as dark-matter particle. A gas of such 
${\cal G}$s would `freeze-out' before the electroweak era as CDM, whose present mean density is predicted 
by this model (via Casimir-effect data on earth) as, roughly, $100{\rm Mev}/{\rm cm}^3$.

\vskip 1cm
\no
\newline
\no
\newline
\vskip 3cm\no
[file: DMm2]
\vfill
\no

\end{titlepage}
\vfill
\eject

\no
\section {Introduction}
 
Following Schwarzschild's solution in 1915, Einstein's deep  interest on `point'
singularities relating to particle aspects and geodesic motion
was expanded and carried all the way into the 50s
with singular solutions (one is
the Papapetrou-Majumdar), but also with point-like particles of
finite radius $r_o$ and the
{\em geon} concept  \cite{e}. The latter had not been realized 
up to now with any proper example, namely any exact non-singular 
electrovacuum which is asymptotically flat, stable, etc \cite{ab}. 
In quite lesser adversity, the  Taub-NUT vacuum was formulated 
within the span of two decades \cite{t}, treated as a cosmological
model at the time. Remarkably, it still is the only explicitly
available solution in the entire class of (even locally) 
$S^3\times\IR$ {\em non-singular} Ricci-flat manifolds\footnote
{The Taub and NUT spaces have boundaries (`null squashed-$S^3$ Misner
bridges' in Taub-NUT) which
are not mathematical singularities (as in a black-hole), but
{\em physical} ones (with non-singular Riemann, etc).}
\cite{rs}. Meanwhile, the introduction of  topological geons
and concern for  stability  re-focused interest
back to singular models and toward the quantum-mechanical 
properties of geon black holes and
Reissner-Nordstr\"om types of Taub-NUT solutions \cite{gp}. 
These developments, combined with the outlook for supersymmetry
under recent LHC results \cite{as}, have motivated us in
uncovering  ${\cal G}$ as the first example of a {\em proper} geon model (GM),
actually a 2-parameter family of GMs.
A specific member therein, the ${\cal G}_{\rm dmp}$,
is proposed as dark-matter particle (DMP) for a new type
of cold dark matter (CDM) models, alternative to the current 
plethora of supersymmetric ones \cite{v}.

As a solitonic pp-wave, ${\cal G}$
propagates along a sourceless  primordial electromagnetic (EM) field $F$ 
of an Einstein-Maxwell electrovacuum. 
As a manifold, ${\cal G}$ is a non-singular left-$SU(2)$ invariant
Bianchi-type IX,  with an extra
Killing vector for axial rotations
as the only survivor of right-$SU(2)$ invariance. ${\cal G}$ has a Taub-NUT 
form of metric and  line element with basic scale 
${\rm L_o}$ \cite{rs}, expressible in terms of the left-$SU(2)$ invariant
1-forms $\ell^i$ (with $\ell^3=\cos\theta d\phi+d\psi$, 
$(\ell^1)^2+(\ell^2)^2=d\Omega^2$ etc,
as parametrized by the $\theta, \phi, \psi$ angles
on $S^3$) and duals $L_i$,  in
\be
d\ell^i\!=\!-\frac{1}{2}\epsilon^i_{jk}\ell^j\wedge\ell^k,\;\;\;\;
 [L_j,L_k]=\epsilon^i_{jk}L_i,
\label{l}
\ee
\be
ds^2&=&-{\rm L_o^2}\left(g\ell^3+2du\right)\ell^3
+r^2d\Omega^2,\;\;\;
\;\left(d\Omega^2=d\theta^2+\sin^2\theta d\phi^2\right).
\label{tm}
\ee
$r=r(u)$, $g=g(u)$ are functions
of the scaleless null $u\in\IR$, used (for now) in lieu
of a $t$-time coordinate.
For the $L_i$, $\partial _u$ vectors we read off (\ref{tm})
that $L_3\cdot\partial _u\!=\!-{\rm L_o^{\,2}}$,
$\partial _u^{\,2}=0$, etc, 
and we'll also find $(L_3)^2=-{\rm L_o^2}/P$ for any $P>0$ constant.
In this geometry, our Lagrangian
\be
{\cal L}=\frac{1}{\kappa^2}R-\frac{1}{4}F^2
\label{l}
\ee
complies with our premise
that any point-like sources of  ($m_s,Q$)
mass, electric charge,  etc, can only
emerge {\em a posteriori}  in ${\cal G}$, effectively or otherwise, if at all\footnote 
{This geometric (`pure-marble') approach will
also avert the formation of a 
mathematical singularity.} \cite{rs}. 

\no
\section {${\cal G}$ compared and contrasted to Taub-NUT}

${\cal G}$ and Taub-NUT share the generic
line element (\ref{tm}) and its isometries.
They also share invariance under $u\rightarrow -u$ 
reflections or $u\rightarrow u+u_o$ translations, and,
as it will turn out, the entire $r=r(u)$ function itself, including
its minimum $r=r_o$ value as a NUT charge\footnote
{This concurrence allows the first
direct identification of a NUT charge:
via the upcoming (\ref{r}) we find $2r_o=\kappa Q$ as the
geometric mean of the fundamental couplings, 
thus a Planck-scale length if $Q^2=1/137$.}.
However, $g(u)$ in (\ref{tm}) turns out to have
the mentioned  $g=1/P$ constant value {\em everywhere} 
in ${\cal G}$, whereas the same $1/P$  constant is approached 
only asymptotically by $g(u)$  in Taub-NUT. 
This result is at the basis of several (and some very
profound) differences in ${\cal G}$ vs  Taub-NUT.
In addition to the $P>0$ parameter, ${\cal G}$ also has the electric charge
$Q\neq 0$ as a second free parameter
(with no counterpart in the also 2-parameter Taub-NUT solution).
And  in addition to the metric scale
${\rm L_o}$ and the NUT charge $r_o$
as a pair of two arbitrary lengths, ${\cal G}$  
also carries the gravitational coupling $\kappa^2=8\pi G_N$
as a third scale. Expectedly, the ${\rm L_o},r_o$ pair
can be expressed in terms of $\kappa$ via the $P$,$Q$ pair, actually as
${\rm L_o}=\kappa\sqrt{P}Q$ and $2r_o=\kappa Q$.
In the case of ${\cal G}_{\rm dmp}$ in particular,
the range of the $r\geq r_o$  radius of the spacelike 
dimensions of the $S^3$  in ${\cal G}$
covers the realistic and fully diversified  hierarchy 
(without supersymmetry!) 
of  {\em four}  fundamental scales as
\be
r_o\;<<\;r_{\rm ew}\sim{\rm L_o}\;<<\;r_{\rm cl}\;<<\;r_H=H_o^{-1}\,,
\;\;\;\;\;\;r_{\rm cl}^3=\frac{3m_{\rm dmp}}{4\pi\rho_{\rm dm}}\,.
\label{ro}
\ee
Here, $r_o$ is Planck-scale,
$r_{\rm ew}\sim{\rm L_o}$ scales
the effective mass in $(m_s, Q)$,  
$r_{\rm cl}$ is the classical mean-free-path of DMPs
(of mass $m_{\rm dmp}$ each)  in
a fluid of mean density $\rho_{\rm dm}$, and
$r_ H=H_o^{-1}$ is the Hubble radius
in a cosmological model filled with that fluid. 
Other profound differences in ${\cal G}$ vs Taub-NUT
involve $L_3$ as an {\em always} timelike vector. Thus,
with no Taub sector or Misner bridges, ${\cal G}$
{\em must} be the $C^\infty$ union (at $u=0$) 
of two  NUT-like pieces ${\cal S_\pm}$
(in $u\rightarrow-u$ symmetry to each other),
as a ${\cal G}={\cal S_-}\vee{\cal S_+}$. 
The ${\cal S_\pm}$ are solitonic pp-waves along
the null $\partial_u$  wave-vector, which satisfies the $D\partial_u=0$ condition,
with ${\cal S_-}$ propagating backwards in time (towards $u=0$) as an antisoliton.
Crucial is the presence  in ${\cal G}$ (along a
spacelike $L_1\times L_2$ radial  direction)
of the $E_C=Q/r^2$  electric field, a {\em primordial}  one,
because no actual charge $Q$ can exist anywhere in ${\cal G}$,
but also  a non-singular one, because there  exists no `$r=0$' origin either.
The $r$ coordinate cannot cover 
${\cal G}$ globally.
This can be seen in terms of the double-valued
$u=\pm|u(r)|$, obtained from the single-valued $r=r(u)$ in the
upcoming (\ref{r}). However, the $u=\pm|u(r)|$  branches (defined over the same
$r\geq r_o$ range with  a spurious singularity at $r=r_o$)
cover quite  elegantly the ${\cal S_\pm}$ submanifolds  in
${\cal G}={\cal S_-}\!\vee{\cal S_+}$ as single-valued functions
over $u\geq 0$, $u\leq 0$, respectively. 
Although practically 
identical to a Taub-NUT already at $r\sim r_{\rm ew}$ and beyond,  
${\cal G}$ cannot reduce
to Taub-NUT or to any  Ricci-flat manifold, or to a 
$Q=0$  limit. The $r_o=0$ limit is also forbidden, so
${\cal G}$ cannot reduce to a singular  limit
(e.g., Kerr-newmann or Reissner-Nordstr\"om) either. 

\no
\section {The solution for ${\cal G}$}

\begin{figure}
\begin{center}
\includegraphics[width=12cm]{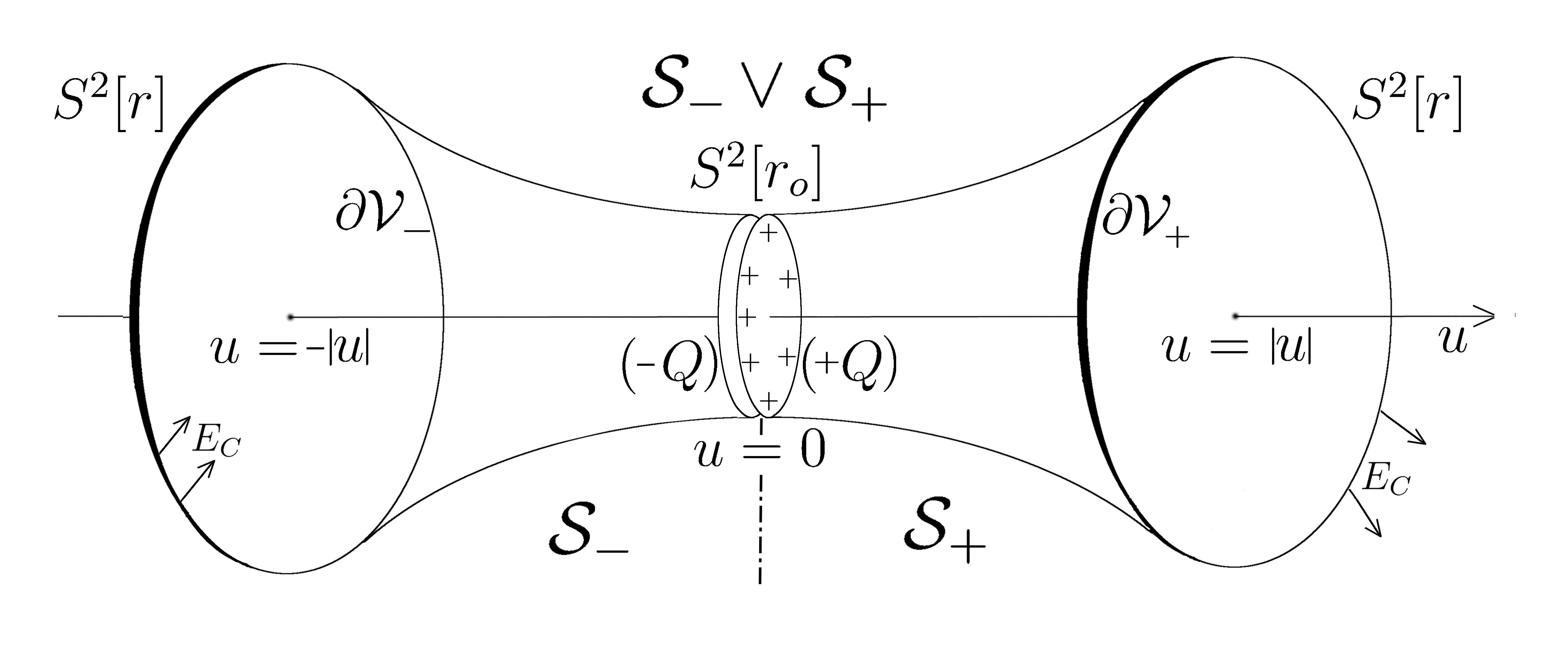}
\caption{A pair of $S^2[r]$ as  (i)  ${\partial{\cal V_\mp}}$
boundaries of a volume ${\cal V}$
in ${\cal S_-}\!\vee{\cal S_+}$ by (\ref{div}), (ii) sections 
of $S^3$ (with its {\small TL} radius suppressed) in
${\cal S_-}$, ${\cal S_+}$ {\em disjointed}, hence 
with $\mp Q$ charge  on the respective
$S^2[r_o]$ boundary (physical singularity)  at $u=0$.
$|E_C|=|Q|/r^2$ everywhere in ${\cal G}$.}
\label{fig}\;
\end{center} 
\end{figure}

Non-holonomic Cartan frames $\theta^\alpha$ (with dual $\Theta_\alpha$)
oriented in the line element (\ref{tm}) of ${\cal G}$ as 
\be
\theta^0={\rm L_o}\left(\sqrt{P}du+\frac{1}{\sqrt{P}}\ell^3\right),\;\;\;
\theta^1=r\ell^1,\;\;\;\theta^2=r\ell^2,\;\;\;
\theta^3={\rm L_o}\sqrt{P}du,
\label{cf}
\ee
\be
\Theta_0=\frac{\sqrt{P}}{\rm L_o}L_3,\;\;\;
\Theta_1=\frac{1}{r}L_1,\;\;\;\Theta_2=\frac{1}{r}L_2,\;\;\;
\Theta_3=\frac{\sqrt{P}}{\rm L_o}\left(\frac{1}{P}\partial_u-L_3\right),
\label{cF}
\ee
give a manifest locally Minkowski
$ds^2=\eta_{\alpha\beta}\theta^\alpha\theta^\beta$, with 
$\eta_{\alpha\beta}={\rm diag}[-1,1,1,1]$. The $D\partial_u=0$ condition
is then verified via
$D\theta^\alpha:=d\theta^\alpha+
\Gamma^\alpha_{\;\beta}\wedge\theta^\beta=0$, which also 
supplies the Christofell $\Gamma^\alpha_{\;\beta}=
\Gamma^\alpha_{\;\beta\gamma}\theta^\gamma$. 
The curvature ${\cal R}^\alpha_{\;\beta}
=1/2R^\alpha_{\;\beta\gamma\delta}\theta^\gamma\wedge\theta^\delta
=d\Gamma^\alpha_{\;\beta}+
\Gamma^\alpha_{\;\gamma}\wedge\Gamma^\gamma_{\;\beta}$
is calculable from 
\be
-\Gamma^0_{\,12}=\Gamma^0_{\,21}=\frac{\rm L_o}{2\sqrt{P}r^2},\;\;
\Gamma^1_{\;20}=-\frac{\sqrt{P}}{\rm L_o}-\frac{\rm L_o}{2\sqrt{P}r^2},\;\;
\Gamma^1_{\;23}=\frac{\sqrt{P}}{\rm L_o},\;\;
\Gamma^1_{\;31}=\Gamma^2_{\;32}=\frac{r'}{\rm L_o\sqrt{P}r},
\label{chris}
\ee
with a prime for $d/du$. Ricci's $R_{\alpha\beta}=R^\gamma_{\;\alpha\gamma\beta}$
follows from the {\em contractible}  Riemann's
\be
R^0_{\,101}=R^0_{\,202}=-\frac{\rm L_o^2}{4Pr^4},\;
R^1_{\,212}=-\frac{1}{\rm L_o^2P}\left(\frac{r'}{r}\right)^{\!2}
+\frac{4Pr^2+3{\rm L_o^2}}{4Pr^4},\;
R^1_{\,313}=R^2_{\,323}=-\frac{1}{{\rm L_o^2}P}\frac{r''}{r},
\label{riemann}
\ee
while Weyl's $W^\alpha_{\;\beta\gamma\delta}$ 
(with just one independent
non-identically-zero component) vanishes as $O(r^{-3})$.
For compatibility with the geometry, 
$F=-E\theta^0\wedge\theta^3+B\theta^1\wedge\theta^2$
is the general  EM field in ${\cal G}$. Then, via the
sourceless Maxwell equations  $dF=d\ast F=0$ which give
 \be
E=-\frac{Q}{r^2}\cos\rho,\;
B=\frac{Q}{r^2}\sin\rho,\;\;\;\;\;r^2d\rho={\rm L_o^2}du,\;\;
\;\;\;\;\;E^2+B^2=\frac{Q^2}{r^4}=:E_C^{\;2}\,,
\label{ems}
\ee
Einstein's $\kappa^{-2}R_{\alpha\beta}=T_{\alpha\beta}^{\rm (em)}=
\frac{1}{2}E_C^{\;2}\,{\rm diag}[1,1,1,-1]$ give
the non-singular general solution
\be
r^2=r_o^2+{\rm L_o^2}Pu^2&,&\;
g=\frac{1}{P}>0\,,\;\;\;\;\;\;\;\;\;\;\;
r_o^2=\frac{\rm L_o^2}{4P}=\left(\frac{\kappa Q}{2}\right)^2\,,
\label{r}
\\
E=E_C-\frac{2r_o^2Q}{r^4}&,&
\;B=-\frac{2r_oQ\sqrt{r^2-r_o^2}}{r^4}\approx
-\frac{{\sf m}_s}{r^3}\,.
\label{em}
\ee
Thus, $F$ includes the Coulomb-like $E_C=Q/r^2$ field and
analogues of an electric quadrapole
and a magnetic-dipole  moment ${\sf m}_s=\kappa Q^2$, with duality
rotations allowed.
To first clarify  the crucial presence of $E_C$ 
in ${\cal G}$ (in spite of  $d\ast F=0$), we apply the divergence theorem
over a 3D volume ${\cal V}$, which includes the $u=0$ locus within 
its $\partial{\cal V}$ boundary  in ${\cal S_-}\vee{\cal S_+}$, as
\be
0=\int_{\cal V}\!d\ast F=\int_{\partial{\cal V}}\!\ast F=
\int_{\partial{\cal V_+}}\!\!\ast F+
\int_{\partial{\cal V_-}}\!\!\ast F=\left[4\pi Q\right]_{\cal S_+}
+\left[-4\pi Q\right]_{\cal S_-}\,.
\label{div}
\ee
As depicted in Fig.\ref{fig}, 
the ${\partial{\cal V_\pm}}$ parts of $\partial{\cal V}$
are round-$S^2$ sections which can be viewed as
evolving from $S^2[r_o]$ 
at $u=0$ to  large absolute values of $\pm|u|$.
The minus sign in the second square bracket in (\ref{div}) is due to the
backwards-in-time propagation of ${\cal S_-}$ as an antisoliton.
The overall null result holds for {\em any} ${\cal V}$, so
there is no $Q$ to be trapped anywhere  in ${\cal S_-}\!\vee{\cal S_+}$, and
the electric flux is not interrupted through any $S^2$ section, notably through
$S^2[r_o]$. Thus, $E_C=Q/r^2$  
must indeed be identified as a sourceless
primordial field  in ${\cal G}={\cal S_-}\!\vee{\cal S_+}$, which
strongly resembles a particle-antiparticle $\bar{p}p$ neutral 
bound state (e.g., a positronium). Such states are typically unstable,
in contrast to ${\cal G}$ which inherits its basic aspects 
(including its stability) as an exact solution
via symmetry and dynamics directly from the Lagrangian (\ref{l}).
To reconcile the
Coulomb-like $E_C=Q/r^2$ in ${\cal G}$ with the
electric-dipole field of a $\bar{p}p$ state, one must
resort to the backwards-in-time propagation of ${\cal S_-}$ as a
submanifold\footnote
{This  is a geometric analogue for antisolitons, after
the  QFT standard for the propagation of antiparticles.}. 
In terms of global  $t$-time as in (\ref{t}), 
for every $(+|t|,\theta,\phi,\psi)$ point
(where $E_C=Q/r^2$)  in ${\cal S_+}$, there is the equally-present
$(-|t|,\theta,\phi,\psi)$ point (where $E_C=-Q/r^2$)  in ${\cal S_-}$, so
the $\pm Q/r^2$  contributions
cancel-out  exactly
at $(t,\theta,\phi,\psi)$ as a {\em single event} in this
${\cal G}$-with-pointwise-identifications-manifold,
now renamed ${\cal G}$.
However, as we'll see, the same $t$ can also
involve differing values of  $r$ in $\pm Q/r^2$, hence 
cancellations which are {\em almost} exact. 
The `corrections' are insignificant 
for sufficiently large $r$, but they do leave
as dominant overall contribution an {\em effective} 
electric-dipole moment ${\sf p}$. This, for ${\cal G}_{\rm dmp}$,
is a minuscule ${\sf p}_{\rm dmp}\sim 4\pi\kappa Q^2=4\pi{\sf m}_s$.

The ${\cal S_\pm}$ can be separated as
independent  geodesically-incomplete  manifolds,
coverable globally by $r$, as seen.
The  boundaries  at $r=r_o$, now exposed as   spurious loci of 
the formerly common $S^2[r_o]$ section (junction) at $u=0$,
become {\em physical} $S^2$ singularities of radius $r=r_o$.
These, under the new initial-value setting (case ii in Fig.\ref{fig}),
must now carry actual $\pm Q$ electric charge,
distributed homogeneously on the respective $S^2[r_o]$ of ${\cal S_\pm}$.

\no
\section {Asymptotic aspects
 and stability of  ${\cal G}$, ${\cal S_\pm}$}

With $R^\mu_{\;\nu\rho\sigma}$
vanishing by (\ref{riemann})  as $O(r^{-2})$ or faster,
the ${\cal G}$, ${\cal S_\pm}$ 
are asymptotically locally flat  manifolds.
They cannot radiate, due to 
$W^\alpha_{\;\beta\gamma\delta}\sim O(r^{-3})$ 
and to stationarity with vorticity
\be
\omega=\ast(v\wedge dv)=\frac{2r_o}{r^2}\,\theta^3\,,
\label{o}
\ee
as measured by an observer
with 4-velocity $V$, dual of $v=\theta^0$  or
$v=dt+{\rm L_o}/{\sqrt{P}}\cos\theta d\phi$  
from the upcoming (\ref{m}).
${\cal G}$, ${\cal S_\pm}$ also
carry {\em calculable} effective mass ($\sim\!m_s$), spin $S$,
dipole moments, etc,
well-defined and measurable by our inertial observer.
With
\be
x^{\mu}=(x^0,x^i),\;\;\;\;\;
x^0=t={\rm L_o}\left(\frac{1}{\sqrt{P}}\psi+{\sqrt{P}}u\right)\;\;\;\;\;\;
\left[{\rm mod}\;8\pi r_o\right],
\label{t}
\ee
namely with  holonomic time\footnote
{The $\psi\in[0,4\pi)$ angle  in (\ref{t})
allows non-trivial homotopy and Planck-scale loops  in the $t$ coordinate.}  
 $t\in\IR$  and Cartesian $x^i$
coordinates  in (\ref{tm}), manifest general covariance
can be traded for what is usually 
taken as a perturbation over Minkowski's  $M_{\!o}^4$
to produce $M^4$ with its metric split as $\eta_{\mu\nu}+h_{\mu\nu}$.
Here, however, we can have $M^4={\cal G}$  {\em exactly} 
(but at a price, as we'll see shortly), if (\ref{tm}) is re-written
(with ${\rm L_o}\sqrt{P}u=\pm\sqrt{r^2-r_o^2}$) as
\be
ds^2&=&-\left(dt+\frac{\rm L_o}{\sqrt{P}}\cos\theta d\phi\right)^2
+\frac{r^2}{r^2-r_o^2}(dr)^2
+r^2\left(d\theta^2+\sin^2\theta d\phi^2\right)
\label{m}
\\
&=&\eta_{\mu\nu}dx^\mu dx^\nu
+\frac{r_o^2}{r^2-r_o^2}(dr)^2
-4r_o\cos\theta d\phi\left(dt+r_o\cos\theta d\phi\right).
\label{h}
\ee
This general result is particularly important for ${\cal G}_{\rm dmp}$,
whose Planck-scale $r_o$ elevates  the $r\rightarrow r_{\rm ew}$ limit
to {\em asymptotic infinity}, with $h_{\mu\nu}\rightarrow 0$
being  already well-established there.
The result  is fundamental because it actually shows that 
such ${\cal G}$s could have been
created abundantly {\em before} the EW era,
as essentially non-interacting DMPs.
It is also crucial for stability, because those
${\cal G}_{\rm dmp}$ would be unable  to break-up 
{\em after} the EW era, for lack of sufficient
excitation energy\footnote
{Stability would
be enhanced by spontaneous 
symmetry breaking for a fixed $\psi=\psi_s$ angle,
which would also lift the homotopy in (\ref{t}).
Time loops at Planck scale  would still exist, but no-longer as geodesic curves.}. 
The price for these deeper findings has been the
loss of manifest left-$SU(2)$ invariance  in (\ref{m}), due to the
absorption of the $\psi$ angle in the definition of $t$ in (\ref{t}). 
In particular, the surviving 
$\theta,\phi$ angles  in (\ref{h}) 
could (and here they do) hinder the calculation of
the mass and spin ($m_s$,$S_s$) parameters
\cite{w} (p.165 ff). Nevertheless, 
an estimate for $m_s$, with a less reliable one for $S_s$,  is possible
via $T_{00}^{\rm (em)}$ and $\omega$ from  (\ref{o}) as
\be
m_s \approx\frac{\sqrt{2}\pi^2 Q^2}{\rm L_o}\sim m_{\rm dmp}\,,\;
S_s \sim\frac{\sqrt{2}\pi^2 Q^2}{\sqrt{P}}\,,\;\;
{\sf m}_s=\kappa Q^2,\;\;
{\sf p}_{\rm dmp}\sim 4\pi\kappa Q^2,\;\;
\rho_{\rm dm}=\frac{3m_{\rm dmp}}{4\pi r_{\rm cl}^3}\,.
\ 
\label{ms}
\ee
These results, which also include the 
mean density $\rho_{\rm dmp}$ of the ${\cal G}$-fluid 
(a  CDM candidate) from (\ref{ro}), etc, hold for ${\cal S_\pm}$,
${\cal G}$, etc, as the case may be.

\no
\section {Conclusions}

By our last results in (\ref{ms}),
the disjointed ${\cal S_\pm}$ carry mass $m_s$
and charge $\pm Q$. ${\cal G}_{\rm dmp}$
as a neutral bound state has mass $m_{\rm dmp}$ 
and no electric charge, but 
it does carry the  effective
dipole moment ${\sf p}_{\rm dmp}$.
The process of uncovering the latter
attribute is actually 
administering a new type of pp-wave {\em superpositions}
(via the point-wise-identifications in section 3, etc), by which the generic
${\cal G}={\cal S_-}\!\vee{\cal S_+}$ manifold has been formed
out of its solitonic constituents.
Analogous superpositions could
be made  possibly even in Taub-NUT, but certainly in a class of generalized
${\cal S_-}\!\vee{\cal S_+}$ GMs. In fact, there exists
a major class of Einstein-Maxwell-Yang-Mills GMs, whose
4D parameter-space includes 
${\cal G}$ as a 2D subclass therein  \cite{b}.

${\cal G}$ is a long-overdue proper GM and,
as a first application thereof, ${\cal G}_{\rm dmp}$ 
founds a CDM model based on those results. 
As a 2-parameter family, ${\cal G}$ can involve drastically different physics from 
the particular ${\cal G}_{\rm dmp}$. This  is closely related to
the range of the hierarchy in (\ref{ro}) and
depends crucially on the 
virtually free-to-choose NUT charge $2r_o=\kappa Q$.
For ${\cal G}_{\rm dmp}$, the NUT charge is a Planck-scale $2r_o$.
It is conceivable, however, that $r_o$ could  be
of a very different scale, e.g., electroweak, or all
the way up to scales of  astrophysical interest.
$Q$  is involved in ${\cal G}_{\rm dmp}$  as the $Q^2=1/137$ coupling
(there is no actual electric charge in any ${\cal G}$)
but, in general, there is no restriction on its value. This is also
important for ${\cal S_\pm}$ as independent manifolds,
which have emerged as a
hardly-expected but nonetheless fundamental result.
As such, the ${\cal S_\pm}$ offer mathematically
non-singular alternatives to the  Kerr-Newman
solution (Reissner-Nordstr\"om is excluded because
there is no $\omega=0$ limit in our case).
Thus, the ${\cal S_\pm}$ upgrade the primitive notion of a
singular point-charge $Q$ to a
rigorous mathematically non-singular solitonic model.  
Also upgraded in the present case is the formerly-singular formation of the
$Q/r^2$ field-lines  and geodesics, in the
sense that the geodesic incompletenesses 
of ${\cal S_\pm}$ is essentially curable,
as actually realized in ${\cal G}$. In any ${\cal G}$ or the ${\cal S_\pm}$
manifolds, the local `03'  Minkowski planes
spanned by $\partial_u$ and $L_3$
carry $F$, $\omega$, $S$, etc. 
They also carry a small subset of 4-velocity  
vectors $V$, which, expectedly
as in Taub-NUT, generate
splitting geodesics and closed timelike loops. This, however,
is a better-accepted reality 
in the case of  ${\cal G}_{\rm dmp}$, because  it 
takes place at Planck scale, as we'll briefly discuss later on.

Another fundamental result for the ${\cal G}$, ${\cal S_\pm}$ configurations
is their mentioned stability under perturbations,  EM or gravitational, 
provided no point-like sources are introduced `by hand'
(c.f. footnote$^ 2$). In the presence
of any spacetime singularity, in particular of a  mathematical one
(e.g., a Kerr-Newman), with which the
${\cal G}$, ${\cal S_\pm}$ would certainly interact, they would fully comply with 
the generally established dynamics and conservation laws
near and within the respective  horizons.
We can now review how ${\cal G}_{\rm dmp}$  
could indeed provide an alternative to
supersymmetric candidates as a geon (solitonic) DMP.
For such an identification, and having established  neutrality and stability,
we must also confirm the virtual absence of
interactions. This is obviously true for the gravitational
(with asymptotic infinity already established at the
EW scale, as seen via (\ref{h})),
as well as for the strong  interaction (which would
be totally absent).
The stability of ${\cal G}$
as an exact solution and the actual value of the effective ${\sf p}_{\rm dmp}$
in (\ref{ms}) will shape any exchange of photons 
between ${\cal G}_{\rm dmp}$ themselves and
between a ${\cal G}_{\rm dmp}$ and baryonic matter. 
As we'll see in an order-of-magnitude calculation, 
the dominant actual EM-field
content in a ${\cal G}_{\rm dmp}$,
essentially from ${\sf m}_s$ and the
effective ${\sf p}_{\rm dmp}$ dipole moments, involves 
much weaker values than standard-model  estimates
for, e.g., a neutrino. Moreover,
the values for ${\sf m}_s,{\sf p}_{\rm dmp}$ are 
smaller by a factor of at least $10^{-10}$
compared to the astrophysically-tolerable limits 
for such EM  moments, so the
interaction between 
${\cal G}_{\rm dmp}$ and baryonic matter
must be accordingly weaker.
The conclusion here is that, in addition to its importance
as the first explicit example of a proper GM, ${\cal G}$
deserves attention because it may have indeed 
been  realized in nature as 
a ${\cal G}_{\rm dmp}$ configuration.

If so, a fluid of these DMPs 
at asymptotic infinity to each-other would
`freeze-out'  before the EW era as
a {\em viable} CDM configuration \cite{v}. Such  DMPs 
could even be trapped today (via induced polarization)
as a mono-layer between equipotential
plates (as local `12' planes) {\em less} 
than $r_{\rm cl}$ apart in a Casimir-effect setting. The trapping
would grow stronger as the gap between the plates
decreases to a sufficiently-smaller-than-$r_{\rm cl}$ value and as
the trapped ${\cal G}_{\rm dmp}$ start orienting
themselves and contributing to force along the `03' plane. 
To fix the ($P,Q)$ set of independent parameters
for order-of-magnitude estimates, we try as $(m_s,Q)$ 
values similar to those of an electron. Then, (\ref{r}), (\ref{ms}) give
$r_o\sim(5\times 10^{19}{\rm Gev})^{-1}
\sim 10^{-33}{\rm cm}$ (a Planck-scale radius),
$ r_{\rm ew}\sim{\rm L_o}\sim 10^{-14}{\rm cm}$
(which decreases if  $m_s$ increases toward typical EW values),
while $\sqrt{P}\sim 10^{19}$. Finally, for the expectedly
tiny dipole moments we find 
${\sf m}_s\sim 10^{-23}\mu_{\rm B}$ with
a likewise tiny ${\sf p}_{\rm dmp}$ in (\ref{ms}).
This result is $10^{-11}$ times smaller than what 
is set by the best astrophysical limits
for a neutrino magnetic dipole moment,  
namely ${\sf m}_{\nu}<\sim 10^{-12}\mu_{\rm B}$.
More  predictions could come via Casimir-effect data on earth.
Thus, from the $r_{\rm cl}\sim 1 {\rm mm}$ gap as a
{\em lower limit} for practically zero-force  
(from ${\cal G}_{\rm dmp}$) 
between the plates, the model predicts all observables
of a  homogeneous  dark-matter universe, including the (roughly)
$\rho_{\rm dm}\sim 100{\rm Mev}/{\rm cm}^3$ density, 
the $r_H= H_o^{-1}\sim 10^{28}{\rm cm}$ Hubble radius, etc.

For the ${\sf p}_{\rm dmp}$ estimate  in (\ref{ms}) we can use the (now
explicitly available) global $t$-time as function of the $(\psi$,$u)$
coordinates  in (\ref{t}) for $t\in\IR$. This
range accommodates the soliton-antisoliton symmetry
under $t\rightarrow-t$ and it does not reduce to $[0,\infty)$ 
in ${\cal G}$ under the point-wise identifications. By these, 
the $\pm Q/r^2$ contributions from the
($\pm|t|,\theta,\phi,\psi$) points
in ${\cal S_\pm}$, respectively, produced
the exact null result at ($t,\theta,\phi,\psi$) as one single
event in ${\cal G}$. However,
we can change the ($\psi$, $u$) pair to
($\psi+\delta\psi$, $u+\delta u$) and yet leave the
given ($t,\theta,\phi,\psi$) intact, 
provided we take $P\delta u=-\delta\psi$ (to secure the same $t$ via (\ref{t}))
and $\delta\psi=4\pi n$, $n\in Z$, 
(to secure the same $\psi$ as an angle on $S^3$). A
$u+\delta u$ value would involve by (\ref{r}) the $r+\delta r$ value
with $\delta r=-8\pi n r_o$ in the $Q/(r+\delta r)^2$ contribution
from ${\cal S_+}$, hence possible non-exact cancellations at the
($t,\theta,\phi,\psi$) event. Thus, an overall contribution may survive
in spite of the $u\rightarrow-u$ symmetry,
because of the presumably random 
involvement of closed loops
near Planck scale and a  thereabout concentration of their effectiveness. The
latter comes from the enormous difference between the 
magnitude of  the $\psi$-term vs 
that of the $u$-term in (\ref{t})
at {\em normal} values of $r$ (that is, not very close to
Planck scale). A  precisely null result for ${\sf p}_{\rm dmp}$ 
might simplify our DMP, but
it is highly unlikely and unnatural to uphold 
the $u\rightarrow-u$ symmetry in a random process at
Planck scale. The order-of-magnitude estimate in (\ref{ms})
as ${\sf p}_{\rm dmp}\sim 8\pi r_oQ$
has thus been used  in lieu of a rigorous result from a
(not-yet available) study  in ${\cal G}$ of such Planck-scale dynamics,
which here seems to feign (or hint) attributes from
an anticipated quantum-gravity environment.

\end{document}